\documentclass[aip,preprint, jmp,showpacs,showkeys]{revtex4-1}

\usepackage[scanall]{psfrag}
\usepackage{graphicx}
\usepackage{amsmath}
\usepackage{dcolumn}
\usepackage{bm}

\usepackage{overpic}
\def \be{\begin{equation}}
\def \ee{\end{equation}}
\def \bf{\mathbf }

\def \CA{\mathcal{C}a}
\begin{document}


\preprint{}

\title{A new prediction of wavelength selection in radial viscous fingering involving normal and tangential stresses.}

\author{Mathias \surname{Nagel}}
  \email{mathias.nagel@epfl.ch}
\author{Fran\c{c}ois \surname{Gallaire}}%
\affiliation{ Laboratory of Fluid Mechanics and Instabilities - EPFL Lausanne, Switzerland
}%




\newcommand{\order}[1]{$\mathcal{O}(#1)$}

\date{\today}

\begin{abstract}

We reconsider the radial Saffman-Taylor instability, when a fluid injected from a point source displaces another fluid with a higher viscosity in a Hele-Shaw cell, where the fluids are confined between two neighboring flat plates. The advancing fluid front is unstable and forms fingers along the circumference.
The so-called Brinkman equations is used to describe the flow field, which also takes into account viscous stresses in the plane and not only viscous stresses due to the confining plates like the Darcy equation. The dispersion relation agrees better with the experimental results than the classical linear stability analysis of radial fingering in Hele-Shaw cells that uses Darcy's law as a model for the fluid motion. 

\end{abstract}


\maketitle 

\section{Introduction}

\begin{figure}[hptb]
 \centering
			\begin{overpic}[width=0.9\textwidth]{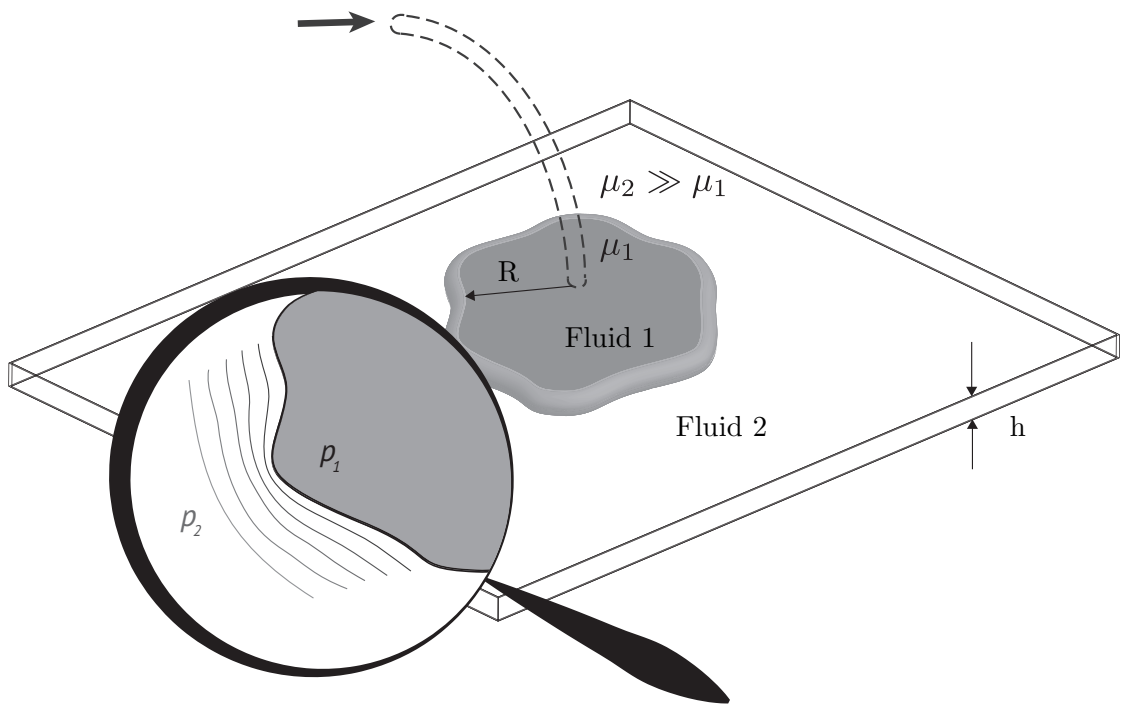}
				\put(60,24){Fluid 2}
				\put(50,32){Fluid 1}
				\put(44,38){R}
				\put(90,24){h}
		\end{overpic}
         \caption{Sketch of an Hele-Shaw cell with an inviscid liquid, injected amid a viscous liquid. A close-up shows schematically the isocontours of the pressure field. }
                    \label{fig:isopressure}
\end{figure}

Viscous fingering, also called Saffman-Taylor\cite{SaffmanTaylor58} instability, is considered as an archetype of pattern forming instability (see Couder 2000\cite{couder2000} for an insightful review). It has also been widely studied in the context of industrial research, such as petroleum extraction in particular. The phenomenon belongs to the broad family of instabilities of growth in Laplacian fields, which includes solidification, aggregation, etc... It was first studied by Saffman and Taylor\cite{SaffmanTaylor58} in 1958, who observed the formation of patterns upon injection of a fluid into a channel filled with another more viscous fluid. Saffman and Taylor\cite{SaffmanTaylor58} studied the formation of fingers in shallow rectangular channels and observed that the formation of fingers was dependent on the ratio of channel height to channel width. The extension to radial geometry, depicted in figure \ref{fig:isopressure}a, dates back to Bataille\cite{Bataille68}, Wilson\cite{Wilson75} and Paterson\cite{Paterson81}.

Viscous and capillary forces govern the mechanism of this instability. The viscosity difference drives the formation of fingers, because emerging fingers enhance the pressure gradients acting in the tip region differentially in each fluid, as illustrated by iso-contours of the pressure field in figure \ref{fig:isopressure}b, where the inner fluid is assumed inviscid and at constant pressure. Since the interface moves with a velocity proportional to the pressure gradient, the feedback loop is closed. In presence of a positive radial gradient of viscosity across the interface, any minute initial displacement will turn unstable, as it induces a pressure gradient difference that further accelerates a protruding finger, which in turn becomes steeper and steeper as it continues to develop. 

This instability is damped by surface tension, which acts to minimize the interface area and opposes the formation of curved fingers. One key parameter that determines the number of fingers is therefore the ratio between viscous effects, represented by the product of the interface velocity $U$ and the viscosity $\mu$, against capillary effects represented by the surface tension $\gamma$. This ratio is called the capillary number $\CA = \frac{U \mu}{\gamma}$.

A second influential parameter is the aspect ratio $k=\sqrt{12}R/h$ (the origin of the factor $\sqrt{12}$ is explained section II). The viscous drag is mainly created by the velocity gradients in the small direction $h$, whereas the capillary forces act on fingers formed along the circumference whose length scales with the radius $R$ (see figure \ref{fig:isopressure}a). 

Considering a flat front separating fluid $1$ to the left from fluid $2$ to the right and propagating along the $x$ coordinate at a velocity $V$ in a rectangular cell of span $b$ (in the direction $y$)  and width $h$ (in the direction $z$), the above physical reasoning has been made rigorous by deriving  a dispersion relation relating the growth-rate $\sigma$ of the instability to the wave number of the disturbance $k$, assuming a normal mode expansion $\exp{(i\alpha x +\sigma t)}$. Chuoke {\it et al.} \cite{Chuoke} could demonstrate that the growth-rate was simply given by:  
\be
\sigma=\frac{\mu_2-\mu_1}{\mu_1+\mu_2} V \alpha -\gamma \frac{\alpha^3}{\mu_1+\mu_2}\frac{b^2}{12}.
\ee
This formula perfectly reflects the physical picture above and shows the existence of a most unstable wavelength, resulting from the competition of a destabilization of the viscosity difference and stabilization by surface tension.
A similar situation holds in the circular injection case, as shown by Wilson\cite{Wilson75}, who determined the wave number of the fastest growing perturbation:
\be
	n = \sqrt{\frac{1}{3}\Biglb( 12\,\CA \frac{R^2}{h^2} +1\Bigrb)},
\ee
where, for brevity, the injected fluid is considered inviscid and unperturbed.
Maxworthy\cite{Maxworthy89} showed experimentally that this prediction works well for low capillary numbers but does not reproduce the behavior at elevated capillary numbers $(\CA >10^{-2})$.

Both Chuoke {\it et al}'s\cite{Chuoke} and Wilson's\cite{Wilson75} predictions are based on the Darcy equation, which is the classical Hele-Shaw limit of the Stokes equation for large aspect ratios. In this limit only viscous terms in the thin direction are retained. The flow in the thin direction is approximated by a Poiseuille profile and the equations can be depth-averaged in the shallow direction. This leads to the Darcy equation, which is a two-dimensional potential flow, where the pressure represents the potential. 
One paradox of this model was pointed out by Dai and Shelley\cite{Shelley93}, who argued that for zero surface tension the fingers would form infinitely sharp cusps.

Paterson\cite{Paterson84} assumed that for zero surface tension flows, for instance with two miscible fluids, the interface is influenced by the full three-dimensional stress tensor. He used a potential flow for which he derived the 3D stress tensor at the fluid-fluid interface and looked for the perturbation wavelength that minimizes the dissipated energy, thereby using the property that low Reynolds number flows minimize the dissipation. He found the finger size $\lambda$ to be solely dependent on the cell height $h$ in the form $\lambda \approx 4.0 h$. This matches quite well the relation Homsy\cite{KimHomsy09} deduced from Maxworthy's results: $\lambda \approx 5.0 h$. A
similar dissipation minimization approach was recently followed by Boudaoud \textit{et al.} \cite{boudaoud}, both for 2D Darcy and 2D Stokes suction flows.

Kim et al.\cite{KimHomsy09} developed a dispersion relation using viscous potential flow where the normal stress boundary condition includes not only pressure but also viscous normal stresses. Their result improved the agreement with Maxworthy's experiment even at capillary numbers of order one. 
In this work the Brinkman equation is used as a model for confined flows. The Brinkman equation retains in-plane stress terms and is therefore of second order, which allows boundary conditions for the normal and tangential velocities at the fluid-fluid interface. Following the same steps as Paterson\cite{Paterson84} in his article for radial fingering using Darcy equations, we are able to show that the finger prediction in the range from low to high capillary numbers is improved. 

In a recent work Logvinov et al.\cite{Smirnov10} used the Brinkman equation to describe the Saffman-Taylor instability of miscible fluids, hence with zero surface tension, in rectangular channels and demonstrate the influence of viscous effects in the flow plane. They found a wavelength dependence that is about $\lambda \approx 2.5 h$, which is still by a factor 2 too small compared to the experiment. Earlier, Fernandez \textit{et al.} \cite{Fernandez} as well as Carl\`es \textit{et al.}\cite{Carles} have also used the Brinkman equation to analyze Rayleigh Taylor instability in Hele-Shaw cells.

The next section \ref{problem} presents first the Brinkman model and thereafter the stability analysis in section \ref{rdd}. In section \ref{results} the results are shown and compare to existing theoretical and experimental results. The paper ends by section \ref{conclusion}, where conclusions and perspectives are drawn.

\section{Mathematical formulation of the problem}
\label{problem}

The flow geometry under consideration is a Hele-Shaw cell, a space confined between two plates, whose distance is generally in the order of a millimeter. The cell is filled with a viscous liquid and in the center is a second fluid of smaller viscosity with a uniformly circular interface. As the injection of more fluid in the center is started the front advances and the instability is triggered. This configuration is schematically depicted in figure \ref{fig:isopressure}a.

The Brinkman equation is used as a model for the fluid motion in this study. Starting from the 3D Stokes equation, it is assumed that the velocity component in the thin direction is zero and the velocity profiles in the plane direction are Poiseuille profiles (see Boos and Thess 1996\cite{Boos} and Gallaire \textit{et al.} 2013\cite{Gallaire} for more details). Under this condition one can integrate in the small direction and obtains depth-averaged equations that combine viscous effects in the plane like in the 2D Stokes equation and the viscous friction due to confinement from top and bottom wall like in the Darcy equation. The parameter related to the aspect ratio that comes from depth averaging is given as $k=\sqrt{12}R/h$, where $R$ is the initial radius and $h$ the height of the cell. This gives $\sqrt{12}$ because the derivative of the parabolic velocity profile in the thin direction is gives $8 R^2/h^2$, since averaged quantities are used one multiplies with the maximum to mean velocity, which is $1.5$ for a parabola. 

In constant flux $Q$ radial injection, the interface evolves as $R=\sqrt{Qt/\pi}$.
For the non-dimensionalisation, using the undisturbed instantaneous radius of the interface $\rho$, the flow rate $Q$, the viscosity of the outer fluid $\mu_2$. This defines the non-dimensional viscosity $\eta_i$ in eq.(\ref{fct:brman}) as $\eta_1 = \mu_1/\mu_2 = \eta$  and $\eta_2 = 1$. This further defines following characteristic velocity and pressure:
\[
	U = \frac{Q}{2 \pi R} \quad \textrm{and} \quad P = \frac{ Q \mu_2}{2\pi R^2}.
\]
The non-dimensional Brinkman equation then reads
\be
	\eta_j (\Delta {\bf u}_j -k^2 {\bf u}_j ) -\nabla p_j = 0 \quad \textrm{with} \quad \nabla \cdot \bf{u}_j = 0.
	\label{fct:brman}
\ee
Herein the velocity components of vector $\mathbf{u}_j$ and pressure $p_j$ are only two-dimensional, as all the operators. The bottom index corresponds to $j= 1$ for the inner fluid and $j=2$ for the outer flow. Given the geometry of the problem it is convenient to write the Brinkman equation in polar coordinates:
		\begin{equation}
			\eta_j \Biglb (\frac{\partial ^2 u_j}{\partial r^2}+ \frac{1}{r}\frac{\partial u_j}{\partial r}+ \frac{1}{r}\frac{\partial ^2 u_j}{\partial \theta ^2}-\frac{u_j}{r^2} -k^2 u_j-\frac{2\partial^2 v_j}{r^2 \partial \theta^2} \Bigrb) - \frac{\partial p_j}{\partial r} = 0
			\label{fct:brinkrad1}
		\end{equation}
		\begin{equation}
			\eta_j \Biglb (\frac{\partial ^2 v_j}{\partial r^2}+ \frac{1}{r}\frac{\partial v_j}{\partial r}+ \frac{1}{r^2}\frac{\partial ^2 v_j}{\partial \theta ^2}-\frac{v_j}{r^2} -k^2 v_j +\frac{2\partial^2 u_j}{r^2 \partial \theta^2} \Bigrb) - \frac{1}{r}\frac{\partial p_j}{\partial \theta} =0.
			\label{fct:brinkrad2}
		\end{equation}
where $u_j$ and $v_j$ denote respectively the radial and azimuthal velocity components. With help of the stream function, $\mathbf{u}_j= (\frac{\partial \psi_j}{r \partial \theta}; -\frac{\partial \psi_j}{\partial r})$, a more compact form is obtained by taking the curl of eq.(\ref{fct:brinkrad1}) and eq.(\ref{fct:brinkrad2}):
\be		
			\Biglb (\frac{1 \partial}{r \partial r}r\frac{\partial}{\partial r}+\frac{\partial ^2}{r^2 \partial \theta ^2} \Bigrb)\Biglb (\frac{1 \partial}{r \partial r}r\frac{\partial}{\partial r}+\frac{\partial ^2}{r^2 \partial \theta ^2} - k^2 \Bigrb)\Psi  = 0.
			\label{fct:stream}
\ee

The kinematic boundary conditions at the interface are equality of normal and tangential velocity, and normal velocity equal to interface velocity:
\be
	\nabla \Psi_1 \cdot \bf{n} =\nabla \Psi_2 \cdot \bf{n} = \dot{\rho}\; , \; \nabla \Psi_1 \cdot \bf{t} = \nabla \Psi_2 \cdot \bf{t}.
\ee

Since the surface stress component in the shallow direction is averaged out, only the radial stress $\sigma_{rr}$ and the in-plane tangential stress $\sigma_{r\theta}$ are matched. The dynamic boundary conditions are continuous tangential stress:
\be
 	\bf{t} \cdot \bar{ \bar \sigma}_1 \cdot \bf{n} =\bf{t} \cdot \bar{ \bar \sigma}_2 \cdot \bf{n},
 	\label{fct:tanstress}
\ee

And jump in normal stresses according to Laplace's law:
\be
	\bf{n} \cdot \bar{ \bar \sigma}_1 \cdot \bf{n} -\bf{n} \cdot \bar{ \bar \sigma}_2 \cdot \bf{n} =\frac{1}{Ca}\Biglb(\frac{\pi}{4} + \frac{2R}{h} \Bigrb).
	\label{fct:nrmstress}
\ee

The corrected boundary conditions of Park and Homsy\cite{ParkHomsy84a} for an advancing menisc of a non wetting fluid is applied, where the in-plane curvature is corrected by $\pi/4$





\section{Linear perturbation theory}\label{rdd}

The stability of the unperturbed interface is analyzed. The unperturbed interface position is $\rho(t_0) =1$ and the base flow for the radially evolving liquid is simply obtained using the incompressibility equation. Since the base flow depends only on the radial direction the pressure can be integrated using eq. (\ref{fct:brinkrad1}),  which gives:
		\[
		\nonumber
			u_{0j} = \frac{1}{r},\quad v_0 = 0 \quad \textrm{and} \quad p_{0j} = -\eta_j k^2 \ln(r) +P_{0j}.
		\]
The solution for the pressure field depends on the constant $P_{01}-P_{02}=1/Ca \Bigl( \frac{\pi}{4} + \frac{2 R}{h} \Bigr)$, which matches the jump in normal stress according to Laplace Law, due to surface tension at the interface. 

	Assuming a perturbation expansion of the non-dimensional interface position $\rho = \rho_0 +\epsilon f(t)e^{i\,n\, \theta}$ and $\Psi_j=\Psi_{0j}+\epsilon \Psi_{1j}$, with a time dependent amplitude $f(t)$ and normal mode Ansatz of wave number $n$. 
	

The Brinkman equation being linear, the general solution for eq. (\ref{fct:stream}) is used. \begin{equation}
		\Psi_{11} = i \Biglb (a_n \frac{I_n(k r)}{I_n(k)} + b_n\, r^n\Bigrb) e^{i\, n\theta},
		\label{fct:innerpsi}
\end{equation}	
\begin{equation}
		\Psi_{12} = i \Biglb (c_n \frac{K_n(k r)}{K_n(k)} + d_n\, r^{-n} \Bigrb) e^{i\,n\theta}.
		\label{fct:outerpsi}
\end{equation}	

The general solution is build of powers of $r$ and modified Bessel functions of first and second kind, $I_n$ and $K_n$. The parameters $a_n$, $b_n$, $c_n$ and $d_n$ have to be determined from the boundary conditions at the interface. The structure of eq. (\ref{fct:stream}) shows that the classical solutions of the Laplace equation are also solutions of the Brinkman equation. Applying flattened boundary conditions at the interface that consider the first order Taylor expansions around the original interface position. 
 
\textbf{Impermeability:} The interface moves with the inner radial velocity such that $\dot \rho + \epsilon \dot f \, e^{i\,n\theta} = u_{01}+\epsilon u_{11}$, where the dot symbol denotes differentiation with respect to time. This yields
	\begin{equation}
		\dot{f} e^{i n \theta} = \frac{\partial u_{01}}{\partial r}  f e^{i n \theta} -\frac{1}{r}\frac{\partial \Psi_{11}}{\partial \theta} + \mathcal{O}(\epsilon) \Big|_{r=1},
	\end{equation}
	which further leads to:	
	\be
		n(a_n+b_n)-f-\dot f = 0. 
		\label{fct:bc1}
	\ee

	\textbf{Continuity of radial velocity:} Matching the outer velocity to the inner velocity at the interface, $u_{11}=u_{12}$:
	\begin{equation}
		\frac{\partial \Psi_{11}}{\partial \theta} = \frac{\partial \Psi_{12}}{\partial \theta} \Big|_{r=1}
	\end{equation}
leads to
		\be
		a_n+b_n-c_n-d_n = 0.
		\label{fct:bc2}
	\ee
	
	\textbf{Continuity of angular velocity:}
	The equality of tangential velocities on the flattened interface, $v_{11} = v_{12}$:
	\begin{equation}
		\frac{\partial \Psi_{11}}{\partial r} = \frac{\partial \Psi_{12}}{\partial r} \Big|_{r=1},
	\end{equation}		
	leading to:
	\be
		a_n\,k \frac{I_{n+1}(k)}{I_n(k)} + c_n\,k \frac{K_{n+1}(k)}{K_n(k)} +n( a_n + b_n -c_n+ d_n) = 0.
	\label{fct:bc3}
	\ee
	
	\textbf{Continuity of tangential stress:} 
	To ensure continuity of the tangential stress at the interface eq. (\ref{fct:tanstress}):
	\[
			\eta \Biglb(-\frac{\partial^2}{r^2\partial \theta^2}\Psi_{11} +r\frac{\partial}{\partial r}\Biglb(\frac{\partial}{r \partial r}\Psi_{11}\Bigrb) \Bigrb)
	\]
	\begin{equation}		
		 = \Biglb(-\frac{\partial^2}{r^2\partial \theta^2}\Psi_{12} +r\frac{\partial}{\partial r}\Biglb(\frac{\partial}{r \partial r}\Psi_{12}\Bigrb) \Bigrb) \Big|_{r=1}
	\end{equation}
	
	Which simplifies to:	
	\[
		\eta \Biglb( 2 a_n\, n^2 + 2 b_n \, n^2 + a_n\,k^2 -2a_n\,k \frac{I_{n+1}(k)}{I_n(k)}-2a_n\,n-2b_n\,n \Bigrb) 
	\]
	\be
		-2c_n\,n^2-2d_n\,n^2-c_n\,k^2-2c_n\,k\frac{K_{n+1}(k)}{K_n(k)}+2c_n\,n-2d_n\,n = 0.
	\label{fct:bc4}	
	\ee
	
	\textbf{Laplace Law:}	
	 The jump of the normal stress balance is more involved since it requires to recover the pressure, which had been left out in the calculation. Integrating eq. (\ref{fct:brinkrad2}) for $\theta$ relates the pressure perturbations to the general solutions.
	 \[
	 	p_{1j} = \eta_j r \int_0^\theta \Biglb(\frac{\partial^3 \Psi_{1j}}{\partial r^3}+\frac{\partial^2 \Psi_{1j}}{r \partial r^2}
	 \] 
	\begin{equation}
			-\Biglb( \frac{n^2}{r^2}+k^2 +\frac{1}{r^2} \Bigrb) \frac{\partial \Psi_{1j}}{\partial r} +\frac{2 n^2}{r^3}\Psi_{1j}  \Bigrb) d\theta
	\end{equation}
which simplifies into	\[
		\nonumber
		p_{11} = - \eta b_n\, r^n\,k^2 e^{in\theta}, \quad \textrm{and} \quad p_{12} = d_n\, r^{-n} k^2 e^{in\theta}.
	\]
The radial stress balance in order $\epsilon$ for the flattened interface according to eq. (\ref{fct:nrmstress}):
	\[
	\nonumber
		p_{11}+\frac{\partial p_{01}}{\partial r} f e^{i n \theta}+2 \eta \frac{\partial}{\partial r}\Biglb(\frac{\partial}{r \partial \theta}\Psi_{11}\Bigrb)-2 \eta \frac{\partial^2 u_0}{\partial r^2} f e^{i n \theta}
	\]
	\[
		-p_{12}-\frac{\partial p_{02}}{\partial r} f e^{i n \theta} -2  \frac{\partial}{\partial r}\Biglb(\frac{\partial}{r \partial \theta}\Psi_{12}\Bigrb) +2 \frac{\partial^2 u_0}{\partial r^2} f e^{i n \theta}
	\]
	\begin{equation}
		= -\frac{\pi }{4\, Ca}\Biglb(1-n^2 \Bigrb) f e^{i n \theta} \Big|_{r=1}
	\end{equation}	
	
	Recalling that the factor $\pi/4$ in front of $1/Ca$ stems from a corrected boundary condition by Park and Homsy\cite{ParkHomsy84a}. Inserting the expressions for velocity and pressure results in:
	\[
		\eta \Biglb( k^2 f(t) + b_n k^2 - 2 a_n n -2 b_n n+2 a_n n k\frac{I_{n+1}(k)}{I_n(k)} + 2 a_n n^2  
	\]	
	\[
		+ 2 b_n n^2 +4 f(t) \Bigrb)+ d_n k^2- f(t) k^2+2c_n n +2d_n n 
	\]
	\be
		+ 2 c_n n k \frac{K_{n+1}(k)}{K_n(k)}  -2 c_n n^2 +2 d_n n^2 -4 f(t) = \frac{\pi}{4 \,Ca} f (1-n^2).
	\label{fct:bc5}
	\ee	
	
	Using the boundary condition equations (\ref{fct:bc1}), (\ref{fct:bc2}),  (\ref{fct:bc3}),  (\ref{fct:bc4}) and  (\ref{fct:bc5}) the unknowns $f, a_n, b_n, c_n, d_n$ have to solve an eigenvalue problem, with $\dot{f}$ as a Eigenvalue. Assuming that the growth of the instability is much faster than the evolution of the base flow, it is assured that the aspect ratio $k$ does not change with time, which leads to the possibility to express the function $f(t) = f_0 exp(\sigma t)$. In this case $\dot f/f = \sigma$, where $\sigma$ is the growth rate.

\section{Results Growth rates for linear theory}
\label{results}
\begin{figure}[htp]
					\psfrag{ylabel}{$\frac{n_{max}}{k} \sqrt{\frac{12}{Ca}}$}
                    \center{\includegraphics[width=1.0\textwidth]{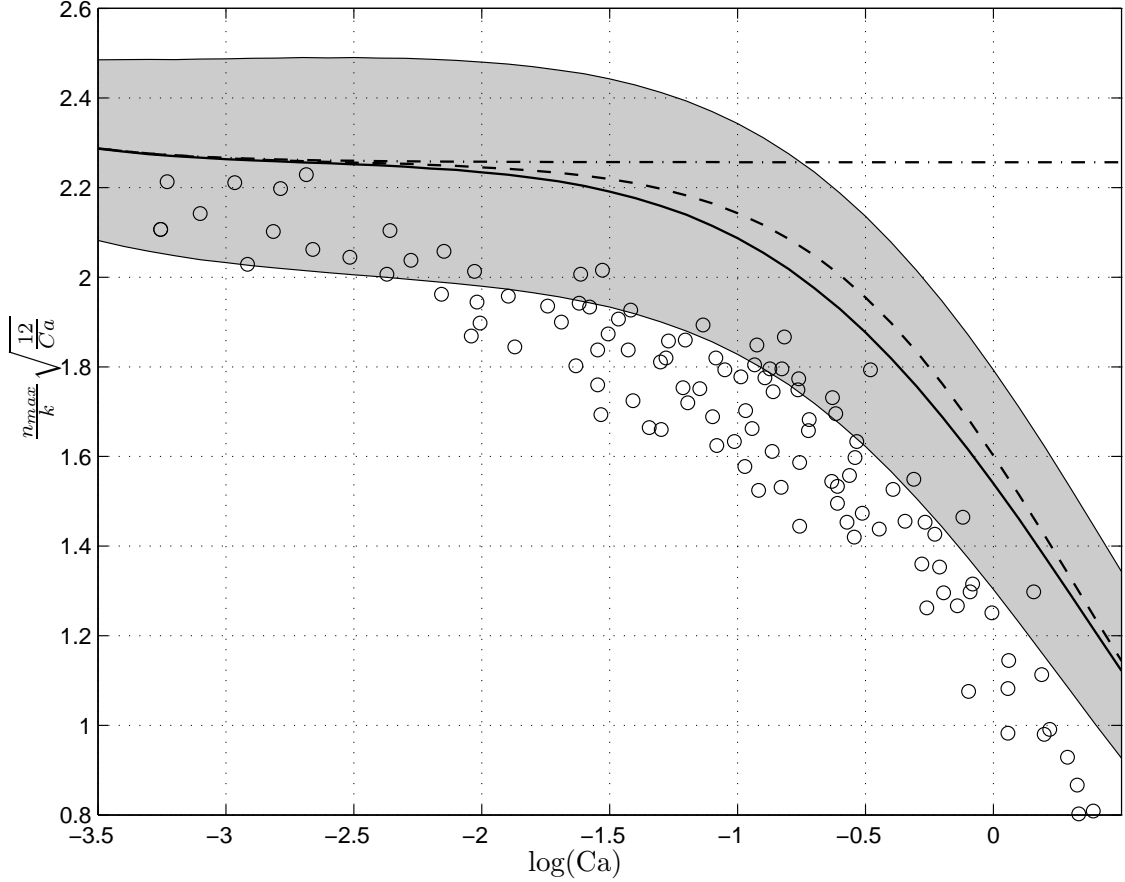}}                
                    \caption{Most unstable modified wave number against logarithm of capillary number. Plotted results are: $- \cdot  -$ Potential flow, - - - Viscous potential flow and --- Brinkman model with aspect ratio $k=200$. Maxworthy's experimental results are also shown, indicated by $\circ$ symbols. The grey shaded region includes all wave numbers of the Brinkman model, where the growth rate is within 1 \% of the maximal growth rate.}
                    \label{pic:paraB}
\end{figure}

	One is interested in the wave number $n$ with the highest growth rate $\sigma$ for given parameters $k$ and $Ca$. In figure \ref{pic:paraB} the most unstable wave numbers plotted in the same way as done by Maxworthy are shown, including the results from our stability analysis. Maxworthy chose to plot the results for a modified wave number being: 
      		\[
      		\nonumber
      		A_{max} = \sqrt{\frac{12}{\textrm{Ca}}}\,\frac{ n_{max}}{k}, 
      		\]
      	because then Paterson's result is approximately represented as a straight line. With the corrected boundary conditions of Park and Homsy this line should be constantly $2.26$ for all capillary numbers. Maxworthy showed, as the capillary number approaches one the observed number of fingers becomes much smaller than predicted by Paterson. Our findings are represented by the full line and show an improved agreement.

\begin{figure}[htp]
					\psfrag{insert}[tc][Bc][1][64]{Increasing Ca}
                    \center{\includegraphics[width=0.8\textwidth]{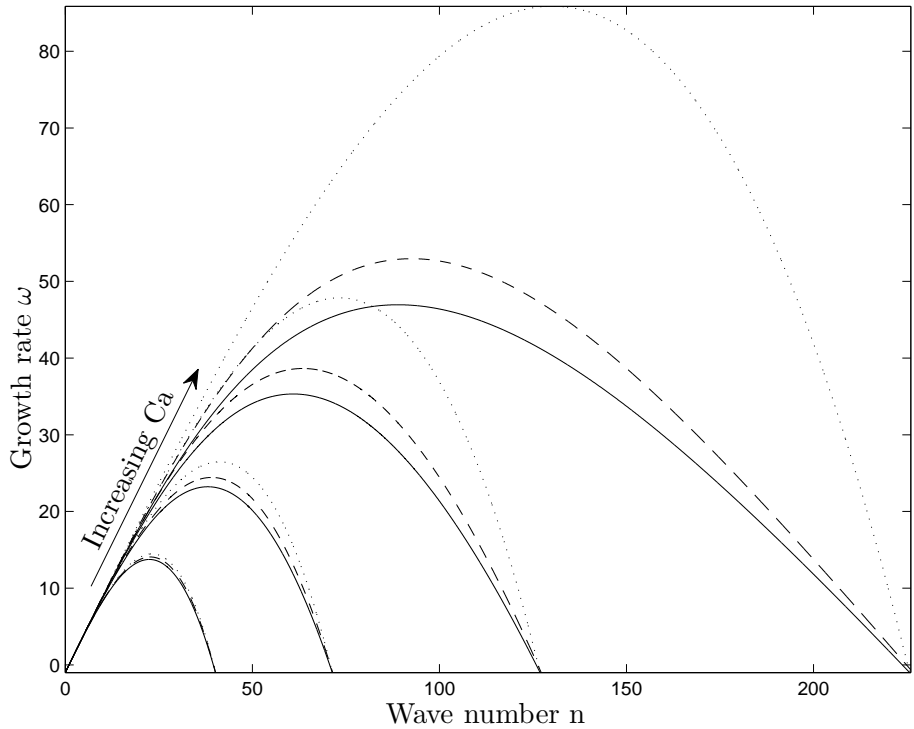}} 
                    \caption{Dispersion relations using different models: $\cdots$ Potential flow, - - - Viscous potential flow and --- Brinkman model. The aspect ratio is fixed to k = 400, while the capillary number changes between $0.032, 0.1, 0.32$ and $1.0$\,.  }
                    \label{pic:paraA}
    \end{figure}

      	Maxworthy's data is spread, which can be explained by the relatively flat summit of the dispersion relation, as seen figure \ref{pic:paraA}, which does not cause a sharp wavelength selection. Displaying also possible fingers that have a growth rate of no less than $1\%$ of the maximal growth rate. The spread takes an almost symmetric shape about the maximal growth rate. 
	
	\begin{figure}[htp]
					\psfrag{ylabel}[b][b]{Finger-number to aspect-ratio: $n_{max}/k$}
                    \center{\includegraphics[width=0.8\textwidth]{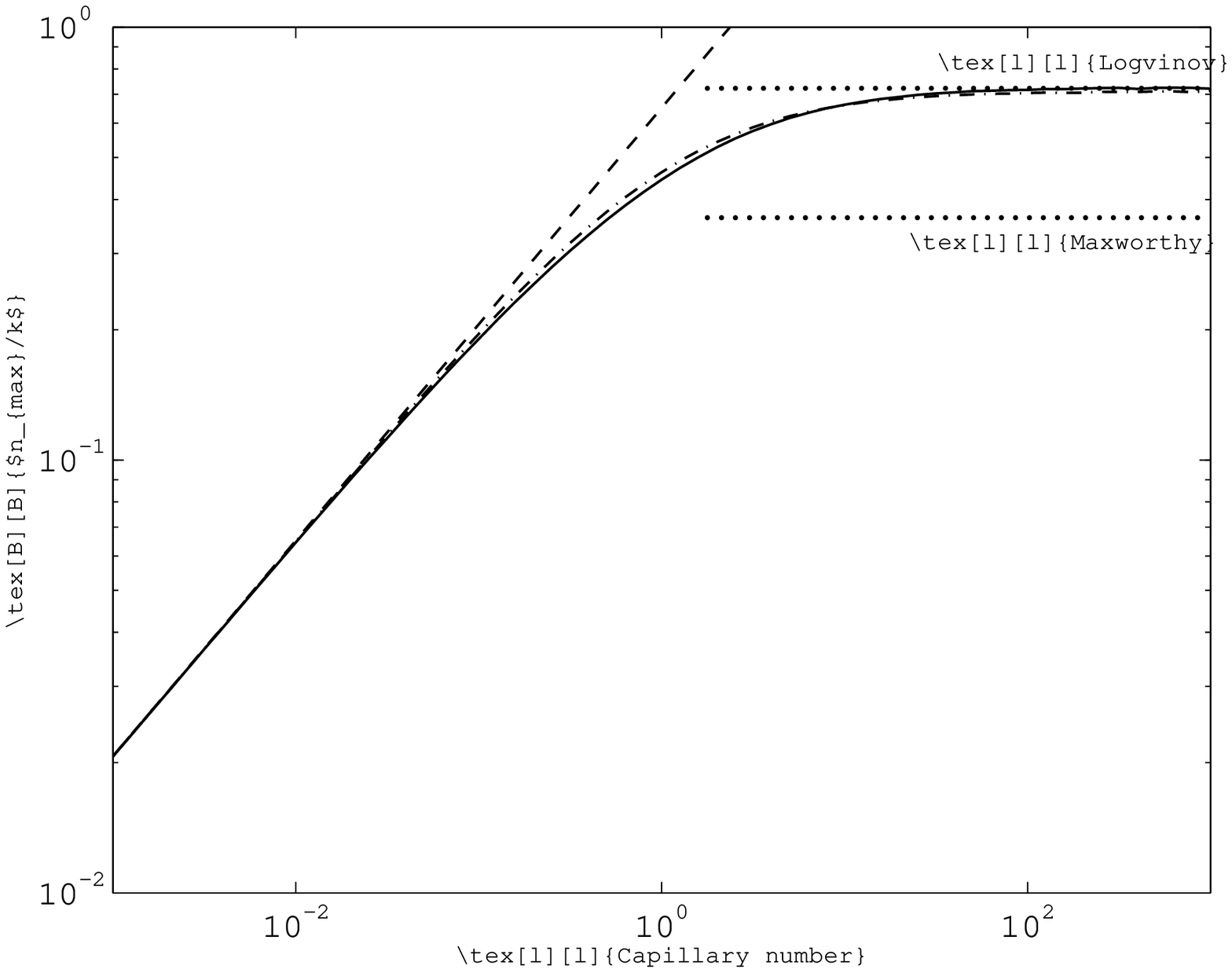}}
                    \caption{ Most unstable finger number as a function of capillary number: $- - -$ Potential flow, $- \cdot -$ Viscous potential flow and --- Brinkman model. Dotted lines show the experimental result of Maxworthy for low surface tension as deduced by Homsy\citep{KimHomsy09} and the theoretical result of Logvinov et al. who used the Brinkman model for zero surface tension fingering of a flat interface with aspect ratio $k = 200$. For higher capillary number the number of fingers becomes proportional to the gap width.}
                    \label{pic:paraC}
    	\end{figure}
	
	The curves vary slightly for low aspect ratio $k< 10^2$. But in experimental applications the aspect ratio was chosen sufficiently large, $k > 10^2$ so the graph collapses to a single curve for all values of $k$ in the range of interest.  	
	
	The dispersion relation that is obtained from the Eigenvalue problem is for brevity given only for the case where the inner viscosity is zero (the general expressions are given in the appendix). The growth rate $\sigma$ becomes: 
	
	\begin{equation}
		\sigma =  \frac{A(k,n)+\frac{\pi}{4}\textrm{Ca}^{-1}\, B(k,n)}{C(k,n)}.
		\label{fct:omegaabc}
	\end{equation}		
	In this form it can be easily seen that the growth rate depends on a parameter pair, where $A(k,n)$ weights the flow rate and $B(k,n)$ weights the influence of surface tension.
	The parametric functions are:
	\[
		A(k,n) =  k^4n-k^4-8n^2-8n^4+4k^2n-8k^2n^2
	\]
	\begin{equation}
		+\frac{K_{n+1}(k)}{K_n(k)} (-2k^3+2k^3n+4kn+4kn^3),
	\end{equation}
	\begin{equation}
		B(k,n) =k^2n-k^2n^3+4n^4-4n^2
		+\frac{K_{n+1}(k)}{K_n(k)}  (2kn-2kn^3),
	\end{equation}
	
	\begin{equation}
		C(k,n) = k^4-8n^2+8n^4+4k^2n^2
		+\frac{K_{n+1}(k)}{K_n(k)} (2k^3-4kn^3+4kn).
	\end{equation}
	
	For a better comparison with existing results the growth rate $\sigma$ obtained by Paterson \cite{Paterson81} is decomposed in the same way. Correcting only the influence of the curvature in Paterson's result by factor $\pi/4$, which is done for the corrected boundary condition. 
	
	\begin{equation}
		\dot f= \Biglb( (n-1)-\frac{\pi}{4}\frac{n(n^2-1)}{\textrm{Ca}\, k^2} \Bigrb) f(t),
		\label{fct:Paterson}
	\end{equation}
	
	Reorganizing the equation in the form of eq. (\ref{fct:omegaabc}) gives:
	
	\[
	\nonumber
		\hat A (k,n) = (n-1)k^4,\quad \hat B (k,n) = n(1-n^2) k^2
	\]
	\[
		\quad \textrm{and} \quad	\hat C (k,n) = k^4.
	\]

This rephrasing shows that the results of Wilson\cite{Wilson75} and Paterson\cite{Paterson81} are recovered in the limit of large aspect ratio $k$. However it is apparent that for zero surface tension, where Paterson's pairing of $\hat A(k,n)/\hat C(k,n)$ grows linearly with $n$, our relation has stabilizing terms that grow faster in wave number $n$ than the destabilizing terms. This ensures that for finite aspect ratio $k$, there will always be a finite cut-off wave number $n$.
	
	Plotting in figure \ref{pic:paraA} the growth rate $\sigma$ as a function of the wave number for different capillary numbers at constant aspect ratio $k = 400$ shows two properties in the dispersion relation induced by the Brinkman model: The growth rates are smaller for all wavelengths and the maximum growth rates shift to lower wave numbers as higher modes receive larger damping.

	The number of fingers for zero surface tension has shown to depend only on the gap width or aspect ratio. Thus plotting the number of most unstable finger divided by the aspect ratio $k$ against capillary number in figure \ref{pic:paraC}. For $k> 10^2$ the results approximately collapse on a single curve for all values of $k$. 
For large capillary numbers the prediction tends asymptotically to the result of Logvinov et al.\cite{Smirnov10} who studied zero surface tension flows in rectangular channels using the Brinkman equation. The experimental result, identified from Maxworthy's data, is missed by a factor 2. 
	Not surprisingly the result of Paterson gives a far too high number of fingers.

\section{Conclusion}
\label{conclusion}

A dispersion relation for Saffman-Taylor modeled by the Brinkman equation covers a wide range of capillary numbers and shows a good approximation with experiments. The decreased wave number of the maximal growth is achieved by keeping the in-plane stresses. When the surface tension has only little influence, the in-plane stresses decrease the growth rate for high wave numbers. The high wave numbers create large velocity gradients and thus shear stresses at the interface. Darcy's law can not account for them, in contrast to the Brinkman model, which allows for a cut-off wave number even with zero surface tension.

The finger size for zero surface tension is found to scale like the cell thickness $h$, as found by Logvinov et al.\cite{Smirnov10} who considered a rectangular cell and by Kim \textit{et al.} \cite{KimHomsy09} for radial injection. In the contrast to the latter study, based on viscous potential flow, our solution satisfies continuity of tangential velocity and shear stress perturbations.

The mismatch that is still observed at low capillary numbers might be attributed to the formation of thin films on the advancing front\cite{ParkHomsy84a}. On one side, the deposition of films reduces the gap width and speeds up the fluid. On the other side, the Laplace law should be modified to account for the strong curvature change in the dynamical meniscus region. Whether including these effects does improve the predicted finger number selection is the subject of our current work. 

Another natural perspective is to consider the full nonlinear solution of the radial fingering by the numerical resolution of the two-phase Brinkman equations. While classical tip-splitting phenomena are recovered, our preliminary results point to a strong dependence of the wavelength selection on the initial conditions and background noise, a large statistical deviation is being observed.


\newpage
\section*{Appendix: Full dispersion relation}

The growth rate for wave number $n$ depending $k$, $Ca$ and $\eta$. 

\begin{eqnarray}
\nonumber
\sigma= -( -(4 I_1 \, k^4 Ca \, K_0) - (8 I_1 \, k^3 \, Ca \, K_1) - (32 I_1 \, Ca \, n^2 \, K_0) - 4 I_1 \pi n^2 K_0 + 4 I_1 \pi n^4 K_0 &\\
\nonumber
 - (32 I_1 \, Ca \, n^4 \, K_0) + (16 I_1 \, k^2 \, Ca \, n\, K_0) -(4 I_1 \, k^4 \eta \, Ca \, K_0) - (32 I_1 \, k^2 \, Ca \, n^2 \, K_0)  &
\\
\nonumber
 + (4 I_1 \, k^4 \,n \, Ca \, K_0) + (8 I_1 \, k^3 \, n \, Ca \, K_1) + (16 I_1 \, k \, n \, Ca \, K_1) + I_1 \, k^2 \, n \pi K_0 + 2 I_1 \, k \, n \pi K_1 
\\
\nonumber
       - I_1 \, k^2 \pi \, n^3 \, K_0 - 2 I_1 \, k \pi \, n^3 \, K_1 + (16 I_1 \, k \, Ca \, n^3 \, K_1) - (4 k^4 \, I_0 \eta Ca  \, K_1) + (8 I_1 \, k^3 \eta^2 Ca \, K_1) 
\\
\nonumber
      - (4 k^4 \, I_0 \eta^2 \, Ca \, K_1) +(64 I_1 \eta n^2 \, K_0 \, Ca) + (64 I_1 \eta n^4 \, Ca \, K_0) - (32 I_1 \eta^2 n^2 \, K_0 \, Ca) 
\\
\nonumber
   + 4 \eta I_1 \pi n^2 \,K_0 - 4 \eta I_1 \pi n^4 \, K_0 - (32 \eta^2 n^4 \, Ca \, I_1 \, K_0) -(32 I_1 \, k \, n \eta Ca \, K_1) + (48 I_1 \, k^2 \eta Ca \, n^2 \, K_0) 
\\
\nonumber
   - (4 I_1 \, k^4 \,n \eta Ca \, K_0) - (16 I_1 \, k^3 \, n \eta Ca \, K_1) - (32 I_1 \, k \eta n^3 \, Ca \, K_1) -(32 n^3 \, I_0 \eta k \, Ca \, K_0) 
\\
\nonumber
	- (8 n^2 \, I_0 \eta k^3 \, Ca \, K_0) + (16 n ^2 \, I_0 \eta k^2 \, Ca \, K_1) - (8 n \, I_0 \eta k^3 \, Ca \, K_0) - (16 n^2 \, I_0 \eta^2 k^2 \, Ca \, K_1)
\\
\nonumber
       +(32 n^3 \, I_0 \eta^2 k \, Ca \, K_0) + (4 k^4 I_0 \, n \eta Ca \, K_1) + (16 I_1 \, k \, n \eta^2 Ca \, K_1) - (16 I_1 \, k^2 \eta^2 Ca \, n^2 \, K_0) 
 \\
\nonumber
+ (8 I_1 k^3 n \eta^2 Ca \, K_1) +(8 n^2 \, I_0 \eta^2 k^3 \, Ca \, K_0) + (8 n \, I_0 \eta^2 k^3 \, Ca \, K_0) - (4 k^4 \, I_0 \, n \eta^2 Ca \, K_1) 
\\
\nonumber
	- (16 I_ 1 \, k^2 \eta^2 Ca \, n \, K_0) - 2 \eta I_1 \, k \, n \pi K_1 + 2 \eta I_1 k \pi n^3 \, K_1 - 2 \eta n^2 \, I_0 \pi k \, K_0 + 2 \eta n^4 \, I_0 \pi k \, K_0 
\\
\nonumber
	+ \eta k^2 \, I_0 \, n \pi \, K_1 - \eta k^2 \, I_0 \pi n^3 \, K_1 +(16 \eta^2 n^3 \, Ca \, I_1 \, k \, K_1)) / (4 Ca) 
\\
\nonumber
        /(-I_1 k^4 \eta K_0 - 4 I_1 k^2 n^2 K_0 - 4 I_1 k n K_1 + 4 I_1 k n^3 K_1 - k^4 I_0 \eta K_1 + 2 I_1 k^3 \eta^2 K_1 - k^4 I_0 \eta^2 K_1 
\\
\nonumber
	 - 16 I_1 \eta n^2 K_0 + 16 I_1 \eta n^4 K_0 + 8 I_1 \eta^2 n^2 K_0 - 8 \eta^2 n^4 I_1 K_0 - I_1 k^4 K_0 - 2 I_1 k^3 K_1 + 8 I_1 n^2 K_0 
\\
\nonumber
	- 8 I_1 n^4 K_0 - 4 n I_0 \eta K_1 k^2 + 8 I_1 k n \eta K_1 + 4 I_1 k^2 \eta n^2 K_0 - 8 I_1 k \eta n^3 K_1 - 8 n^3 I_0 \eta k K_0 + 4 n^2 I_0 \eta k^2 K_1 
\\
\nonumber
 - 2 n I_0 \eta k^3 K_0 - 4 n^2 I_0 \eta^2 k^2 K_1 + 8 n^3 I_0 \eta^2 k K_0 - 4 I_1 k n \eta^2 K_1 + 2 n I_0 \eta^2 k^3 K_0 - 4 I_1 k^2 \eta^2 n K_0 
\\
\nonumber
	+ 4 \eta^2 n^3 I_1 k K_1 + 8 n ^2 I_0 \eta k K_0 - 8 n ^2 I_0 \eta^2 k K_0 + 4 n I_0 \eta^2 K_1 k^2 + 4 I_1 k^2 \eta n K_0).
\end{eqnarray}

\normalsize

\end{document}